\documentclass[a4paper,11pt]{article}
\usepackage{pos}

\title{Machine Learning of Thermodynamic Observables in the Presence of Mode Collapse}
\author*[a,b]{Kim A. Nicoli}
\author[a,b]{Christopher J. Anders}
\author[c,d]{Lena Funcke}
\author[e,f]{Tobias Hartung}
\author[g]{Karl Jansen}
\author[a,b]{Pan Kessel}
\author[a,b,h]{Shinichi Nakajima}
\author[i]{Paolo Stornati}

\affiliation[a]{Technische Universit\"{a}t Berlin, Machine Learning Group,
Marchstrasse 23, Berlin,10587, Germany}
\affiliation[b]{Berlin Institute for the Foundation of Learning and Data (BIFOLD),
Technische Universität Berlin, Berlin, Germany}
\affiliation[c]{Center for Theoretical Physics, Co-Design Center for Quantum Advantage, and NSF AI Institute for Artificial Intelligence and Fundamental Interactions,
Massachusetts Institute of Technology,\\ 77 Massachusetts Avenue, Cambridge, MA 02139, USA
}
\affiliation[d]{Perimeter Institute for Theoretical Physics, 31 Caroline Street North, Waterloo, ON N2L 2Y5, Canada}
\affiliation[e]{Computation-Based Science and Technology Research Center, The Cyprus Institute, 20 Kavafi Street, 2121 Nicosia, Cyprus}
\affiliation[f]{Department of Mathematical Sciences University of Bath, Bath, United Kingdom}
\affiliation[g]{Deutsches Elektronen-Synchrotron DESY, Platanenallee 6, 15738 Zeuthen, Germany}
\affiliation[h]{RIKEN Center for AIP, 1-4-1 Nihonbashi, Chuo-ku, Tokyo, Japan}
\affiliation[i]{ICFO, The Barcelona Institute of Science and Technology, Av. Carl Friedrich Gauss 3, 08860 Castelldefels (Barcelona), Spain}

\emailAdd{kim.a.nicoli@tu-berlin.de}
\emailAdd{anders@tu-berlin.de}
\emailAdd{lfuncke@mit.edu}
\emailAdd{tobias.hartung@desy.de}
\emailAdd{karl.jansen@desy.de}
\emailAdd{pan.kessel@tu-berlin.de}
\emailAdd{nakajima@tu-berlin.de }
\emailAdd{paolo.stornati@icfo.eu}

\abstract{Estimating the free energy, as well as other thermodynamic observables, is a key task in lattice field theories. Recently, it has been pointed out that deep generative models can be used in this context \cite{nicoli2021estimation}. Crucially, these models allow for the direct estimation of the free energy at a given point in parameter space. This is in contrast to existing methods based on Markov chains which generically require integration through parameter space. In this contribution, we will review this novel machine-learning-based estimation method. We will in detail discuss the issue of mode collapse and outline mitigation techniques which are particularly suited for applications at finite temperature.\\[2ex]
Preprint number: MIT-CTP/5353}

\FullConference{%
 The 38th International Symposium on Lattice Field Theory, LATTICE2021
  26th-30th July, 2021
  Zoom/Gather@Massachusetts Institute of Technology
}

\begin{document}
\maketitle

\section{Introduction}
Recent works have investigated the use of a particular class of deep generative machine learning models, called normalizing flows, in lattice field theory  \cite{flowsforlattice, flowsforlattice2, flowsforlattice3, del2021efficient, hackett2021flow, nicoli2021estimation, de2021scaling} following similar approaches in quantum chemistry \cite{noe2019boltzmann, wirnsberger2020targeted, wirnsberger2021normalizing} and statistical physics \cite{wu2019solving, nicoli2020asymptotically, nicoli2019comment} (see also \cite{bachtis2021quantum,urban2018reducing,tanaka2017towards, bulusu2021generalization}).
These works are proof-of-principle demonstrations for simple two-dimensional field theories and aim to reduce the integrated autocorrelation time for systems close to criticality by using the flow to generate decorrelated field samples. 

Another important application of flows was recently pointed out in \cite{nicoli2020asymptotically}: they can directly estimate the free energy of a lattice field theory (which can also be accomplished with methods such as tensor
networks, see, e.g., \cite{akiyama2020tensor} and references therein). The free energy is important as it allows to compute the entropy, pressure and the equation of state of the considered physical system. In the case of Quantum Chromodynamics, such thermodynamic observables are of the utmost importance in the physics of the early universe and are probed by heavy ion experiments \cite{busza2018heavy}. 

In the following, we will review this deep-learning-based estimation technique of the free energy and discuss the important issue of mode collapse. To illustrate the mode collapse of the flow in a concrete example, consider a target density with two modes, as is the case for a quantum mechanical particle in a double well potential or scalar $\phi^4$-theory in the broken phase. The latter example will be discussed in detail in Section~\ref{sec:phi4}. For both systems, the theory has a spontaneously broken $\mathbb{Z}_2$ symmetry and thus two modes corresponding to the vacuum expectation values $\pm \Phi$. As we will discuss in Section~\ref{sec:learning}, the training process can however lead to a flow that only approximates one mode of the target density of the lattice field theory and assigns (almost) vanishing probability mass to the other \cite{nicoli2021estimation, hackett2021flow}. This will lead to systematic errors of the free energy estimate which can be difficult to detect. In this contribution, we report on both mitigation and detection techniques for such a mode collapse and demonstrate their effectiveness for the example of two-dimensional scalar $\phi^4$ theory. 

\section{Normalizing Flows}
Let $f:\mathcal{M} \to \mathcal{N}$ be an orientation-preserving diffeomorphism between two orientable $n$-dimensional Riemannian manifolds $\mathcal{M}$ and $\mathcal{N}$. We assume that there is a probability measure $q \,\textrm{d}V$ defined on $\mathcal{M}$ 
where $dV$ is the measure associated with volume form on $\mathcal{M}$ and $q:\mathcal{M} \to \mathbb{R}_+$ is a positive smooth map. In particular, it holds that $\int_\mathcal{M} q \, \textrm{d}V = 1$. 
The push-forward measure $f_*(q \, \textrm{d}V)$ is then a probability measure on $\mathcal{N}$.

In coordinates $(x^i, U)$ on $\mathcal{N}$, the push-forward $f_*(q \, \textrm{d}V)$ takes the form
\begin{equation*}
     q(f^{-1}(x)) \left|f'(f^{-1}(x))\right|^{-1} \, \, \textrm{d} V(f^{-1}(x))  \,,
\end{equation*}
where $\left|f'\right|$ is the determinant of the Jacobian.
In the machine learning literature, one therefore often refers to 
\begin{align}
(f_* q) (x) \equiv  q(f^{-1}(x)) \left|f'(f^{-1}(x))\right|^{-1} \label{eq:logprob}
\end{align}
as the push-forward density of $q$. 

We will be interested in the case $\mathcal{M}=\mathcal{N}=\mathbb{R}^n$ since we will consider real-valued scalar fields. The basic idea of a normalizing flow is to define a family of diffeomorphisms $f_\theta$ with parameters $\theta$. We then adjust these parameters $\theta$ such that the push-forward density $(f_\theta)_*q$ closely approximates a certain target density $p$.

In practice, the diffeomorphism $f_\theta$ is parameterized by a deep neural network. 
Neural networks are composite functions of the form
\begin{align}
m_\theta \colon \mathbb{R}^n &\to \mathbb{R}^n \nonumber \\
z &\mapsto 
    m_\theta(z) \,,
\end{align}
where $ m_\theta(z) = m^{L} \circ \dots \circ m^{1}(z) \label{eq:nn}$ is a composition of layers $m^i$ defined by
\begin{align*}
    m^{i}(z) = \sigma ( W^i z + b^i)\,,
\end{align*}
with weights $W^i\in \mathbb{R}^{n, n}$ and biases $b^i \in \mathbb{R}^n$ being the free parameters of the neural network, i.e. $\theta = \{(W^i, b^i)\}_{i=1}^K$.\footnote{We restrict to all weights and biases being of the same dimensionality since we will be interested in networks that can be used to model invertible maps.}
Furthermore, $\sigma(z)$ is a non-linear function, such as $\sigma(z)=\tanh(z)$, which is applied element-wise to each component of the vector $W^i z+b^i \in \mathbb{R}^n$. A neural network is called deep if the number of layers $L$ is large (although there is no clearly defined threshold).

There are various approaches for parameterizing diffeomorphisms by neural networks. We will restrict to a particularly straightforward approach, called \emph{Non-linear Independent Component Estimation} (NICE), which splits the input $z=(z_u, z_d) \in \mathbb{R}^n$ in two parts $z_u \in \mathbb{R}^{n-k}$ and $z_d \in \mathbb{R}^{k}$ for given $k\in \{1, n -1 \}$. A diffeomorphism is then given by
\begin{align}
f_\theta(z) = 
\begin{bmatrix} 
f_u (z)  \\
f_d (z)
\end{bmatrix}  
= 
\begin{bmatrix}
z_u \\
z_d + m_\theta( z_u )
\end{bmatrix}
  \,,
\end{align}
where $m_\theta$ is a (not necessarily invertible) neural network of the form \eqref{eq:nn}. 
Due to the splitting of the input $z$, this can be easily inverted by
\begin{align*}
\begin{bmatrix} 
z_u  \\
z_d
\end{bmatrix} 
= 
\begin{bmatrix}
f_u \\
f_d - m_\theta( f_u )
\end{bmatrix} \,.
\end{align*}
For the NICE architecture, the determinant of the Jacobian is given by
\begin{align*}
\left| \frac{\partial f_\theta}{\partial x} \right| =
\begin{vmatrix}
\tfrac{\partial f_u}{\partial z_u} & \tfrac{\partial f_u}{\partial z_d}  \\
\tfrac{\partial f_d}{\partial z_u} & \tfrac{\partial f_d}{\partial z_d} \\
\end{vmatrix}
=
\begin{vmatrix}
\mathbb{I} & 0 \\
* & \mathbb{I} \\
\end{vmatrix}
= 1 \,. 
\end{align*}
As a result, the diffeomorphism $f_\theta$ is volume-preserving, i.e. $\left| \frac{\textrm{d}f_\theta}{\textrm{d}z} \right|=1$. In practice, we compose several of these volume-preserving diffeomorphisms. This combination is again a volume-preserving diffeomorphism because these maps form a group under composition.  

A normalizing flow $f_{\theta*} q$ is typically chosen to be a push-forward of a simple base density, e.g. $q=\mathcal{N}(0,1)$. This allows for efficient sampling by first drawing $z \sim \mathcal{N}(0,1)$ and then applying the diffeomorphism $f_\theta$ to the sample $z$, i.e. 
\begin{align}
    f_\theta(z) \sim f_{\theta*} q \,, \label{eq:sampling}
\end{align}
where the push-forward density $f_{\theta *}q$ is given by \eqref{eq:logprob}.

\subsection{Training of the Flow}\label{sec:learning}
A lattice field theory can be described by a probability density of the form
\begin{align}
    p(\phi) = \frac{1}{Z} \exp(-S(\phi)) \,, \label{eq:target}
\end{align}
where $\phi$, $S$ and $Z$ denote the field, its action, and the partition function respectively. 
\\
A similarity measure between two densities $p$ and  $f_{\theta*} q$ is given by the Kullback--Leibler (KL) divergence
\begin{align}
    \textrm{KL}(f_{\theta*} q || p ) = \int \mathcal{D}[\phi] \,  f_{\theta*} q(\phi) \, \log \left( \frac{f_{\theta*} q(\phi)}{p(\phi)} \right) \,.
\end{align}
The KL divergence is non-negative and vanishes if and only if both densities are equal, i.e. $p = q$.\footnote{We restrict to continuous densities here. Otherwise, the densities can have different values on a set of zero measure.} 
We can therefore train the flow to approximate the target density $q$ by minimizing this KL divergence using gradient descent, i.e. $\theta \leftarrow \theta - \nabla_\theta \textrm{KL}(f_{\theta *}q || p)$. 
For this, we observe that the KL divergence can be rewritten as 
\begin{align*}
    \textrm{KL}( f_{\theta *} q || p) = \mathbb{E}_{\phi \sim f_{\theta *} q} \left[ S(\phi) + \log \left(  f_{\theta *} q (\phi) \right)  \right] + \textrm{const.} \,,
\end{align*}
where the last summand contains terms independent of $\theta$ and can thus be ignored for gradient descent.
We now sample from the flow to obtain its Monte-Carlo estimator, i.e.
\begin{align*}
    \textrm{KL}( f_{\theta *} q || p) \approx \frac{1}{N} \sum_{i=1}^N \left[ S(\phi_i) + \log \left(  f_{\theta *} q (\phi_i) \right)  \right] + \textrm{const.} \,, && \phi_i \sim f_{\theta *} q \,.
\end{align*}
The log probability can efficiently be calculated by \eqref{eq:logprob}. Additionally, we can very efficiently sample from the flow by pushing forward samples from the base density, see \eqref{eq:sampling}. However, the training of the flow may yield poor results for a multi-modal target density. This is because the training relies on self-sampling. During training, self-sampling may lead to a collapse of almost all the flow's probability mass to a subset of the modes of the target density $p$. The KL divergence does not penalize this behaviour since the flow does no longer produce samples from the other modes of the target density $p$. We will discuss both detection and mitigation of mode collapse in the next section.

\section{Flow-based Estimation of Free Energy}\label{sec:freeenergy}
A promising application of normalizing flows is estimating the free energy of a lattice field theory at temperature $T$ defined by
\begin{align}
    F = - T \log Z \,,
\end{align}
where $Z$ is the partion function. The temperature is given by $T=\frac{1}{N_T a}$ with lattice spacing $a$ and $N_T$ denoting the number of lattice points along the temporal direction of the lattice.

\subsection{MCMC-based Estimates of Free Energy}\label{sec:mcmc}
Estimating the free energy with MCMC is challenging. To illustrate this fact, we discuss a reweighting procedure \cite{de2001t,philipsen2013qcd} which starts from the observation that the difference in free energies $\Delta F_{e\,b} = F_e - F_b = -T \log (\frac{Z_e}{Z_b})$ between two different points $e$ and $b$ in parameter space can be calculated by
\begin{align}
   \mathbb{E}_{p_b} \left[ \frac{\exp(-S_e)}{\exp(-S_b)} \right]= \frac{1}{Z_b} \int \mathcal{D}[\phi] \, e^{-S_b(\phi)} \,\frac{e^{-S_e(\phi)}}{e^{-S_b(\phi)}} = \frac{Z_e}{Z_b} \,. \label{eq:diffZ}
\end{align}
This expectation value can be estimated by MCMC. If we choose the point $b$ in parameter space such that the free energy $F_b$ can be calculated exactly or approximately, we can obtain the value of the free energy at the point $e$ by $F_e = \Delta F_{e \, b} + F_b$.
\\
In practice, the variance of the estimator \eqref{eq:diffZ} will become prohibitively large if the two distributions $p_b$ and $p_e$ have a small overlap. This can be avoided by choosing intermediate distributions $p_{i_1},\dots p_{i_K}$ such that neighbouring distributions $p_{i_k}$ and $p_{i_{k+1}}$ overlap sufficiently. The free energy difference can then be obtained by
\begin{align}\label{eq:freeenergyerror}
    \Delta F_{e \, b} = \Delta F_{e \, i_K} + \Delta F_{i_K \, i_{K-1}} + \dots + \Delta F_{i_1 \, b} \,.
\end{align}
This comes at the price of an accumulated error of all free energy differences $\Delta F_{i_k i_{k+1}}$. The error therefore crucially depends on all points of the (discretized) trajectory connecting the points $e$ and $b$ in parameter space. 

\subsection{Example: Two-dimensional $\phi^4$ Theory}\label{sec:phi4}
This dependence on the trajectory can lead to serious problems, as we illustrate in a concrete example of the $\phi^4$ theory in two dimensions with the action
\begin{align}
S = \sum_{x \in \Lambda} ( - 2 \kappa \sum_{\hat{\mu}=1}^2 \varphi(x) \varphi(x + \hat{\mu}) + (1 - 2 \lambda)& \varphi(x)^2 
+ \lambda \, \varphi(x)^4 ) \,, \label{eq:action}
\end{align}
where $\kappa$ is the hopping parameter and $\lambda$ denotes the bare coupling. For vanishing hopping parameter $\kappa$, the free energy can be calculated analytically \cite{nicoli2021estimation} and is given by
\begin{align*}
    F(\lambda) = - |\Lambda| \, T \, \ln z(\lambda) \,, 
\end{align*}
where $|\Lambda|$ denotes the number of sites of the lattice $\Lambda$ and
\begin{align*}
z(\lambda) = \sqrt{\frac{1-2\lambda}{4 \lambda}} \, \exp\left(\frac{(1-2 \lambda)^2}{8 \lambda}\right) \, K_{\frac{1}{4}}\left( \frac{(1-2 \lambda)^2}{8 \lambda} \right) \,,
\end{align*}
with $K_{n}$ being the Bessel function of the second kind.

\begin{figure}[tb]
    \centering
    \includegraphics[width=\textwidth]{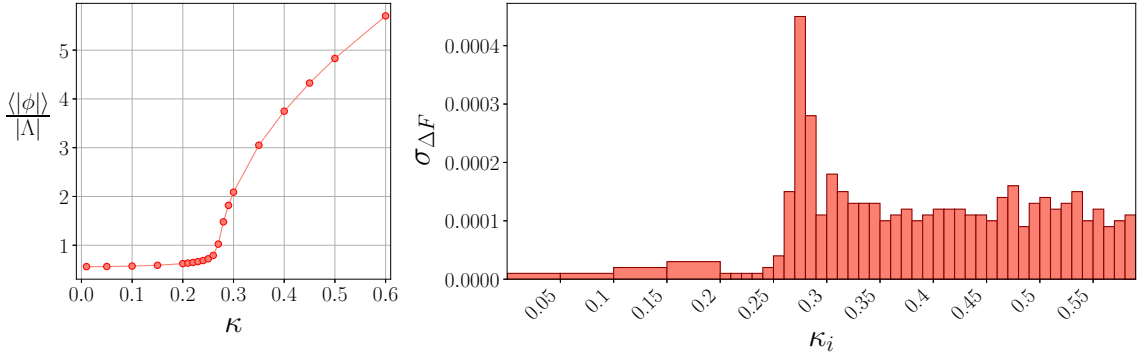}
    \caption{\textbf{Left:} absolute magnetization density $\frac{\langle |\phi| \rangle}{|\Lambda|}$, where $|\Lambda|$ denotes the number of lattice points, as a function of the hopping parameter $\kappa$ for the bare coupling $\lambda=0.022$ on a $16\times 8$ lattice. The values were estimated with an overrelaxed HMC \cite{adler1981over, whitmer1984over, callaway1983lattice, fodor1994overrelaxation}. \textbf{Right:} contributions to the error from the HMC free energy estimate at hopping parameter $\kappa=0.6$ and bare coupling $\lambda=0.022$ for a $16 \times 8$ lattice along the discretized trajectory through parameter space. The width of a bar corresponds to the step size $\Delta \kappa$. The area of a bar shows the error of the corresponding free energy difference $\Delta F$ (calculated as described in \cite{wolff2004monte}), i.e. the total area of all bars is the total error of the free energy estimate $F_e$ at the target value $\kappa_e=0.6$. Every chain is four hundred thousands steps long with an overrelaxation every ten steps. Even for this relatively small lattice, a significant contribution to the overall error comes from the region around the critical value of $\kappa$.}
    \label{fig:absmag}
\end{figure}

As the hopping parameter $\kappa$ is increased, spontaneous breaking of the $\mathbb{Z}_2$-symmetry $\phi \to -\phi$ is observed. This is illustrated in Figure~\ref{fig:absmag}. Now, suppose we want to calculate the free energy with MCMC for parameters in the broken phase, e.g. $\lambda_e=0.022$ and $\kappa_e=0.5$. We can then choose a trajectory through parameter space for which the bare coupling is kept constant, i.e. $\lambda_{i_k}=0.022$, and the initial hopping parameter is $\kappa_b=0$. We then increase the hopping parameter by a step size $\Delta \kappa=0.05$ up to $\kappa=0.2$ and then use a smaller step size $\Delta \kappa=0.01$ in order to ensure sufficient overlap. Crucially, the estimate of the free energy in the broken phase will now suffer from critical slowing down as the corresponding trajectory has to cross the phase transition in order to reach the initial hopping parameter $\kappa_a=0$. This will lead to a significant increase in the statistical error, see Figure~\ref{fig:absmag}.

\subsection{Flow-based Estimators of the Free Energy}
Normalizing flows allow us to directly estimate the free energy $F=-T \log(Z)$ at a given point in parameter space and therefore allow us to avoid critical slowing down in the specific situations discussed in the previous section. This can be seen by observing that we can estimate the partition function $Z$ using a trained flow in two different ways. Firstly, we can use samples from the flow
\begin{align*}
    Z = \mathbb{E}_{\phi \sim f_{\theta *}q} \left[ \frac{e^{-S(\phi)}}{f_{\theta *} q(\phi)} \right] \approx \frac{1}{N} \sum_{i=1}^N \frac{e^{-S(\phi_i)}}{f_{\theta *} q(\phi_i)} \equiv \hat{Z}_q  \,,  && \phi_i \sim f_{\theta *} q \,. 
\end{align*}
Using this definition, we obtain the \emph{reverse estimator} of the free energy by
\begin{align}
    \hat{F}_q = - T \log (\hat{Z}_q) \,. \label{eq:estfirst}
\end{align}
Secondly, one can use samples from the target density $p$
\begin{align*}
    Z^{-1} = \mathbb{E}_{\phi \sim p} \left[ \frac{f_{\theta *}q(\phi)}{e^{-S(\phi)}} \right] \approx \frac{1}{N} \sum_{i=1}^N \frac{f_{\theta *}q(\phi_i)}{e^{-S(\phi_i)}} \equiv \hat{Z}_p^{-1} \,, && \phi_i \sim p\,.
\end{align*}
to obtain the \emph{forward estimator}
\begin{align}
    \hat{F}_p = T \log (\hat{Z}_p^{-1} ) \,. \label{eq:estsecond}
\end{align}
Both estimators have relative strengths and weaknesses. If we are confident that the flow closely approximates the target density $p$, it is advisable to use the reverse estimator \eqref{eq:estsecond} because sampling from the flow is more efficient. However, this estimator may lead to incorrect results if the flow is mode-dropping. In contrast, the forward estimator $\hat{F}_p$ uses samples from $p$ and thus cannot neglect any mode of the target density $p$. If mode-dropping is a risk (for example in the broken phase of the $\phi^4$-theory), one should therefore also use the forward estimator \eqref{eq:estsecond} as a consistency check. 

\section{Numerical Experiments}

\begin{figure}[t]
   \centering
    \includegraphics[width=\textwidth]{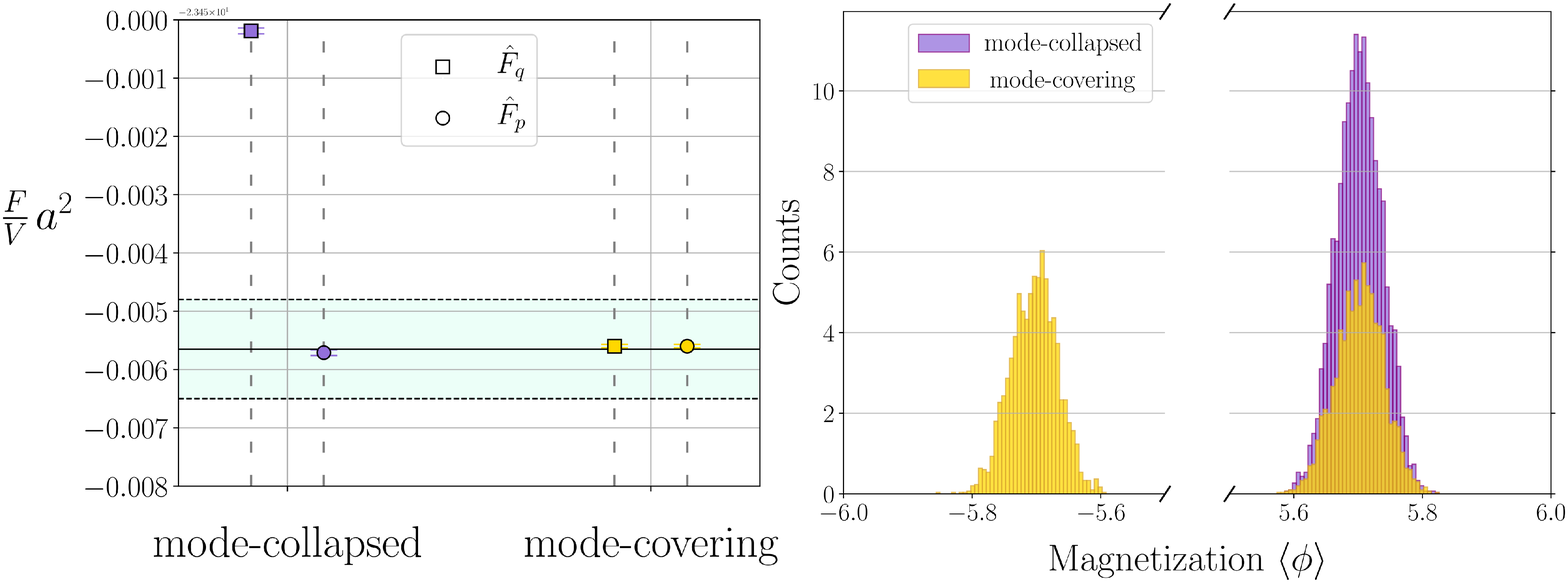}
    \caption{\textbf{Left:} estimation of the free energy using both the forward and reverse estimators. One of the flows has mode-collapsed (purple) while the other is mode-covering (yellow). For the former case, this results in a discrepancy in the free energy prediction of the two estimators while for the latter we obtain consistent results. The MCMC reference values are estimated by the method described in Section~\ref{sec:mcmc}. The mean of MCMC estimate is shown by the solid line while the color band denotes the confidence interval of one standard deviation. As the MCMC algorithm, we again use an HMC with overrelaxation. \textbf{Right:} histogram of the magnetization obtained by direct sampling from both a mode-collapsed and mode-covering flow.}
    \label{fig:modecollapsehist}
\end{figure}

In the following, we will illustrate the difference in using the forward and the reverse variants for the free energy estimation in the presence of mode-dropping.
To this end, we consider two normalizing flows trained for the two-dimensional scalar $\phi^4$-theory for a hopping parameter of $\kappa=0.6$ and a bare coupling of $\lambda=0.022$ on a $N_T\times N_L=16\times 8$ lattice. The theory is thus considered in its broken phase, see Figure~\ref{fig:modecollapsehist}. One of the flows is mode-collapsed on a single mode of the target density $p$, while the other flow covers both modes, as can be seen on the right of Figure~\ref{fig:modecollapsehist}. 
\\
For both flows, we then use the forward estimator $\hat{F}_p$ and reverse estimator $\hat{F}_q$ as defined in \eqref{eq:estsecond} and \eqref{eq:estfirst} respectively. The estimated values for the free energy are visualized in Figure~\ref{fig:modecollapsehist}. For the mode-collapsed flow, we see a clear discrepancy in the prediction while the mode-covering flow leads to consistent values of the forward and reverse estimators. The fact that the forward estimator $\hat{F}_q$ gives the correct result for the mode-collapsed model can heuristically be understood by assuming that the flow is approximately $f_{\theta *} q(x) \approx 2 p(x)$ for the covered mode $\mathcal{M}_1$ and $f_{\theta *} q(x) \approx 0$ for the other mode $\mathcal{M}_2$. This implies that
\begin{align}
    \mathbb{E}_{\phi \sim p} \left[ \frac{f_{\theta *}q(\phi)}{e^{-S(\phi)}} \right] \approx \int_{\mathcal{M}_1} p(\phi) \underbrace{\frac{f_{\theta *}q(\phi)}{e^{-S(\phi)}}}_{\approx \frac{2}{Z}} + \int_{\mathcal{M}_2} p(\phi) \underbrace{\frac{f_{\theta *}q(\phi)}{e^{-S(\phi)}}}_{\approx 0} \approx \frac{2}{Z} \underbrace{\int_{\mathcal{M}_1} p(\phi)}_{\approx \frac{1}{2}} \approx Z^{-1} \,. 
\end{align}
In summary, this experiment clearly illustrates that forward estimation of the free energy is crucial in the presence of mode collapse.

\section{Conclusion}
Deep generative models, in particular normalizing flows, allow for a direct estimation of the free energy. Current normalizing flow architectures are however far from perfect. For example, they are challenging to train in the broken phase (particularly for larger lattices) and can suffer from mode collapse for multi-modal densities. In this contribution, we have briefly outlined how forward estimation of the free energy can help to mitigate this weakness.

\section{Acknowledgements}
K.A.N.\ , C.A.\ , P.K.\ and S.N.\ are funded by the German Ministry for Education and Research as BIFOLD - Berlin Institute for the Foundations of Learning and Data (ref. 01IS18025A and ref 01IS18037A). P.S.\ is supported from Agencia Estatal de Investigación (“Severo Ochoa” Center of Excellence CEX2019-000910-S, Plan National FIDEUA PID2019-106901GB-I00/10.13039 / 501100011033, FPI) ), Fundación Privada Cellex, Fundación Mir-Puig, and from Generalitat de Catalunya (AGAUR Grant No. 2017 SGR 1341, CERCA program). L.F.\ is partially supported by the U.S.\ Department of Energy, Office of Science, National Quantum Information Science Research Centers, Co-design Center for Quantum Advantage (C$^2$QA) under contract number DE-SC0012704, by the DOE QuantiSED Consortium under subcontract number 675352, by the National Science Foundation under Cooperative Agreement PHY-2019786 (The NSF AI Institute for Artificial Intelligence and Fundamental Interactions, http://iaifi.org/), and by the U.S.\ Department of Energy, Office of Science, Office of Nuclear Physics under grant contract numbers DE-SC0011090 and DE-SC0021006. Research at Perimeter Institute is supported in part by the Government of Canada through the Department of Innovation, Science and Industry Canada and by the Province of Ontario through the Ministry of Colleges and Universities.

\bibliographystyle{JHEP}
\bibliography{references}

\end{document}